\documentclass[a4paper,fleqn]{cas-sc}

\usepackage[authoryear,longnamesfirst]{natbib}

\def\tsc#1{\csdef{#1}{\textsc{\lowercase{#1}}\xspace}}
\tsc{WGM}
\tsc{QE}
\tsc{EP}
\tsc{PMS}
\tsc{BEC}
\tsc{DE}
\begin{document}
\let\WriteBookmarks\relax
\def\floatpagepagefraction{1}
\def\textpagefraction{.001}
\shorttitle{Effects of COVID on learning and teaching}
\shortauthors{López Flores et~al.}
%\begin{frontmatter}

\title [mode = title]{Effects of the COVID-19 Pandemic on Learning and Teaching: a Case Study from Higher Education}                      

\author%
[1]
{Nidia Guadalupe López Flores}
\cormark[1]
\ead{nidia20@ru.is}
%\credit{Data analysis, Writing - Original draft preparation}

\author%
[1]
{Anna Sigríður Islind}
\ead{islind@ru.is}
%\credit{Data analysis, Writing - Original draft preparation}

\author%
[1]
{María Óskarsdóttir}
\ead{mariaoskars@ru.is}
%\credit{Data analysis, Writing - Original draft preparation}

\address[1]{Department of Computer Science, Reykjavík University, Menntavegur 1, 102 Reykjavík, Iceland}

\cortext[cor1]{Corresponding author}

\begin{abstract}
In December 2019, the first case of SARS-CoV-2 infection was identified in Wuhan, China. Since that day, COVID-19 has spread worldwide, affecting 153 million people.  Education, as many other sectors, has managed to adapt to the requirements and barriers implied by the impossibility to teach students face-to-face as it was done before. Yet, little is known about the implications of emergency remote teaching (ERT) during the pandemic. This study describes and analyzes the impact of the pandemic on the study patterns of higher education students. The analysis was performed by the integration of three main components: (1) interaction with the learning management system (LMS), (2) Assignment submission rate, and (3) Teachers' perspective. Several variables were created to analyze the study patterns, clicks on different LMS components, usage during the day, week and part of the term, the time span of interaction with the LMS, and grade categories. The results showed significant differences in study patterns depending on the year of study, and the variables reflecting the effect of teachers' changes in the course structure are identified. 
This study outlines the first insights of higher education's new normality, providing important implications for supporting teachers in creating academic material that adequately addresses students' particular needs depending on their year of study, changes in study pattern, and distribution of time and activity through the term.
\end{abstract}

%\includegraphics{figs/grabs.pdf}

%\item Research highlights item 1
%\item Research highlights item 2
%\item Research highlights item 3


\maketitle
% Introduction needs more motivation and context. Maybe we can add some material from the related work. I think it would be good to talk about the extreme situations. We should also present some of the excising COVID+learning research.
% My suggestion
%1.paragraph: Placing the paper in covid
%2.paragraph: Learning/Teaching in extreme situations
%3.paragraph: In this paper... (emphasise our 3 way analysis)
%4.paragraph: Resaerch qestions
%5.paragraph: Details about our anaysis (5 courses, 3 years, ...)
%end with a clear gap in the literature

\section{Introduction}
%1. paragraph  Placing the paper in covid
During 2020 there were many sectors affected by the spread of SARS-CoV-2 virus. COVID-19 was declared a pandemic by the World Health Organization (WHO) on March 11, 2020, \citet{WHO20}. From that day, and to the end of April 2021 when this paper is written, all sectors and industries have had to adapt to, and, in some cases, create a new reality. Education at all levels was one of those sectors. According to the United Nations Educational, Scientific and Cultural Organization (UNESCO), during October 2020 around 1.300 million students were affected by the full or partial lock-downs of educational institutions from pre-primary to tertiary levels of education \citet{UNESCO20}.

%2.paragraph: Learning/Teaching in extreme situations
In the past, crisis and extreme situations have forced the implementation of innovative provisional solutions due to the lack of continuity in the education of the students affected by those situations. In some cases, the impact induced by those situations reached up to 80\% of the local student community. Depending on the particular circumstances of a crisis, some examples of the most common interventions are mail, radio, television broadcasts, and printed material \citep{Creed_Morpeth_2014}. Other examples of educational innovations in extreme situations can be found in refugee camps during armed conflicts; in some cases, the establishment of temporary learning spaces was needed to support primary and secondary education in short to midterm \citep{sinclair2001education}. During the COVID-19 pandemic outbreak, all educational levels, in most of the educational institutions around the world, implemented online teaching and learning, at least to some extent, in order to minimize the impact of school closures \citep{islind2021socio}.

%3.paragraph: In this paper... (emphasise our 3 way analysis)
The changes in distribution and allocation of time due to modifications in course content and teaching methods forced by the pandemic, as well as the differences in the impact on the learning experience of students in different years of their studies, are not yet well documented. There is still a gap in the literature regarding the implications of emergency remote teaching (ERT) during the pandemic. Initial research findings have been obtained, using surveys and questionnaires to document both students' and teachers' views on their first impressions and impact on learning and teaching during the pandemic \citep{benito2021, Aucejo20}. To prevent limitations in survey responses and providing a more extensive landscape, we base our research of the impact and consequences of the pandemic on the study patterns of students on an analysis of three main components. Firstly, the students' interaction with the Learning Management System (LMS) Canvas during the pandemic and how it differs from previous terms; the second component is related to drop out assignment submissions during the pandemic; finally, the last component is the teachers’ point of view, the challenges faced, and the main changes implemented in their teaching due to and during the pandemic. More specifically, in this paper, we thereby analyze and describe the impact that the COVID-19 pandemic had on students’ learning patterns through a data-driven approach, analyzing data from the fall of 2020 and comparing those interactions to the two previous years.

%4.paragraph: Resaerch qestions
Our main research question is: How has the COVID-19 pandemic and those changes implied by it, impacted the study patterns of students in higher education? To answer the question, we collected the click-stream data through the LMS Canvas, the students’ grades and the individual assignment submissions from 2018 to 2020. Additionally, to complement the click-stream data and quantitative analysis, we conducted semi-structured interviews with the teachers to map how the pandemic affected their teaching methods, the challenges faced from their view-point and their expectations about the performance of the students during the pandemic. In addition, the third component is used to understand the patterns identified with the first two components; the interview format allowed the teachers to go into detail about their worries, thoughts and decisions made during the pandemic period. 

%5.paragraph: Details about our anaysis (5 courses, 3 years, ...)
Our analysis encompasses five compulsory undergraduate 12-week courses taught in the autumn terms of 2018, 2019 and 2020, with a total of 956, 1089, and 1113 students enrolled in them, each year, respectively. The courses are part of the Computer Science department at a University in the Nordic countries [University name removed for review purposes]. The Computer Science department was selected because of its high level of online activity in the LMS system before the pandemic hit and during it. The main contribution of this study is an in-depth understanding of the changes in study patterns for higher education students during the pandemic. This contribution allows for the identification of factors in students' learning experience, which can be used to be address similar situations, in future crises. 

%Here I would also talk about changes in behaviour from various aspects and how we measure it, i.e. clicks on different components, clicks on parts of the week, clicks by grade

The structure of the paper is as follows; in Section 2, we discuss the related previous work on learning analytics, extreme situations and the impact of the pandemic on learning and teaching. Section 3 presents the data as well as the methods used for the data analysis. The results obtained from the analysis of each component are presented in Section 4. Finally, Section 5 includes the discussion, conclusion and main contribution of this paper.

%The Elsevier cas-sc class is based on the
%standard article class and supports almost all of the functionality of
%that class. In addition, it features commands and options to format the
%\begin{itemize} \item document style \item baselineskip \item front
%matter \item keywords and MSC codes \item theorems, definitions and
%proofs \item lables of enumerations \item citation style and labeling.
%\end{itemize}

%This class depends on the following packages
%for its proper functioning:

%\begin{enumerate}
%\itemsep=0pt
%\item {natbib.sty} for citation processing;
%\item {geometry.sty} for margin settings;
%\item {fleqn.clo} for left aligned equations;
%\item {graphicx.sty} for graphics inclusion;
%\item {hyperref.sty} optional packages if hyperlinking is
%  required in the document;
%\end{enumerate}  

%All the above packages are part of any
%standard \LaTeX{} installation.
%Therefore, the users need not be
%bothered about downloading any extra packages.

\section{Related Work}

\subsection{Learning Analytics}

Learning analytics is defined as "measurement, collection, analysis and reporting of data about learners and their contexts, for purposes of understanding and optimizing learning and the environments in which it occurs" \citep{LAdef}. According to \citet{Ehe2021} 90\% of the higher education institutions are currently using, or planning to use, learning analytics and see it as a priority for strategic development; while 57\% indicated that they were taking advantage of blended learning by having a range of online resources available, from short courses to fully online degree programs. The survey also showed 95\% of the higher education institutions highlighted digitization as a priority for their approaches to teaching and learning; the revision of teaching methods and the flexibility of provision are perceived as the main benefits of digitally enhanced learning and teaching.

Click-stream and log data have been widely used in learning analytics, with the objective of improving several aspects of learning and teaching for both students and teachers. In these types of studies, click-stream data from LMS and Virtual Learning Environment (VLE) are the main sources of data, which can be complemented with assessment grades, data from other student information systems and with student surveys \citep{TSAI20}. Several fields had been benefiting from its usage; \citet{CASEY17} identified activity patterns of material consumption and usage to detect topics that students find difficult to understand and developed a classification system for early detection of poor performance by students. 

\subsection{Blended and distance learning on extreme situations}

Emergency situations and crises can take many shapes, climatological emergencies and natural disasters like earthquakes, monsoons, tsunamis, and hurricanes; cultural and sociopolitical conflicts, i.e., civil wars, armed conflicts, and  minority ethnic groups limitations and barriers; as well as more complicated and less common situations, like health epidemics, or economic fallout. In response to those situations, many innovative interventions and frameworks have been developed and implemented to maintain the continuity of education provision in crisis situations \citep{Creed_Morpeth_2014, bojovi2020}. 
The concept of Emergency Remote Teaching (ERT) has arisen in need to properly describe the current situation in most of the educational institutions and to clarify its difference from traditional pre-planned online teaching. Online learning considers nine design dimensions: modality, pacing, student-instructor ratio, pedagogy, role of online assessments, instructor role online, student role online, online communication synchrony, and source of feedback \citep{means_bakia_murphy_2014}. However, remote teaching during crisis situations lacks control over some of those dimensions, for instance, modality and student-instructor ratio. While the planning process of a complete pre-planned online course may take from six to nine months before its delivery, a crisis implies a sudden change ranging from a implementation time of a couple of days up to a few weeks in the best scenarios. ERT is defined as a shift in the way the instructional material is delivered, triggered by a crisis. This alternative way, differs from the way that the instructional design was originally planned. ERT involves alternating to fully remote, or online teaching instead of meeting the students in an face-to-face setting. ERT includes the aspiration of switching to face-to-face setting once the crisis has subsided \citep{Emergency20}.

Online and distance learning has many benefits for both learners and instructors; nevertheless, it implies several challenges when it turns out to be ERT. Those challenges can be situated, which indicates that they can vary from one country to another and between institutions, departments and people involved. Furthermore, the challenges can be related to both people and technology. The challenges specifically visible for ERT can for instance be: i) technological challenges, due to internet connection and access to proper devices; ii) pedagogical challenges, digital skills, interactivity level and related to motivation; and iii) social challenges, such as lack of human interaction and family support \citep{ferri_grifoni_guzzo_2020}.

\subsection{Impact of COVID-19 on study patterns}
Several aspects of the way life changed due to the COVID-19 pandemic are still present. Governments around the world implemented restrictive measures to contain the spread of the virus. Consequently, life habits and activities drastically changed in a short period, and people had to adapt to those changes rapidly \cite{covid2020changes}. Remote work and distance education were some of those measures implemented, leading to a massive modification in the usage of technological resources. By April 2020, 94\% of the student population had been affected by the closures of schools and educational institutions. Different approaches were implemented to assist the students in their learning, ranging from online classes in some instances to broadcasting through television channels and radio stations in other instance \citep{unitednations_2021}. Among the recommendations to address the learning crisis during the COVID-19 pandemic, \cite{unitednations_2021} highlights the need to ensure the quality of the data  generated by the students, teachers and learning systems to allow the continuous monitoring of the learning environments. In the higher education sub-sector of education, practically all the institutions managed to move their teaching and learning into online or blended learning, in most cases supported by LMS \citep{Ehe2021}. 

Recent research about the effects of the pandemic on education has focused on varied contexts; \citet{tang_chen_law_wu_lau_guan_he_ho_2021} explored motivation, readiness and self-efficacy during the coronavirus outbreak for sub-degree, undergraduate and postgraduate students; \citet{roy_covelli_2020} examined students reactions and opinions about the emergency switch to online teaching in the middle of the term to provide guidelines for best practices in online pedagogy when returning to the classroom; \citet{benito2021} analyzed the satisfaction levels of students and teachers in three universities; notwithstanding most of them are satisfied with online learning during the pandemic, furthermore they felt the online lectures were less engaging, and the volume of learning was lower than before; \citet{zhang2021measuring} found significant changes in planning and goal setting strategies during school closure by using click-stream data; related to students' performance, \citet{chavez21} analyzed the quality of messages posted and its relation with grades in virtual environments; \citet{xu_wilson_2021} investigated the effect of concept drifts on machine learning models' performance. 

Regarding to online platforms usage and study pattern affectations during the pandemic, \citet{reigal2020} used questionnaires of users in 803 countries to discover changes in the pattern of usage of online platforms. They found a significant increase in the number of users and data traffic generated by a psycho-social assessment platform during the first months of the pandemic compared with the year before.  \citet{Noviyanti20} performed a qualitative and quantitative analysis based on responses to an online survey from high school students, they found significant changes in the activities and learning patterns before and during the pandemic. Among the main changes identified: sleep patterns suffered irregularities and students woke up at a later hour during the pandemic; there was also a significant increase in social media consumption; while in the learning context, a decrease in learning intensity and variety of learning media were present during the pandemic. \citet{Aucejo20} also used online surveys to find differences in study patterns before and after the pandemic. Their results show important variations in study time: about half of the students either increased their study time by more than four hours or decreased it by more than five hours. Study patterns were affected by other factors as well, like changes in the expected graduation year, the need to withdraw from classes, and academic performance; in addition, these effects are heterogeneous and depend on the income, gender, or race of the students. The results also showed that 47\% of the students are less likely to choose online learning in the future rather than on-site learning, due to the bad experiences during the pandemic and the emergency remote teaching. 
Noteworthy, these conclusions are primarily based on surveys and questionnaires; this case study contributes to the gap in the literature regarding the effect of the COVID-19 pandemic in study patterns in higher education students by providing insights into student interactions with the LMS through click-stream data.  

%The package is available at author resources page at Elsevier
%(\url{http://www.elsevier.com/locate/latex}).
%The class may be moved or copied to a place, usually,
%\verb+$TEXMF/tex/latex/elsevier/+, %$%%%%%%%%%%%%%%%%%%%%%%%%%%%%
%or a folder which will be read                   
%by \LaTeX{} during document compilation.  The \TeX{} file
%database needs updation after moving/copying class file.  Usually,
%we use commands like \verb+mktexlsr+ or \verb+texhash+ depending
%upon the distribution and operating system.

\section{Methodology}

%The author names and affiliations could be formatted in two ways:
%\begin{enumerate}[(1)]
%\item Group the authors per affiliation.
%\item Use footnotes to indicate the affiliations.
%\end{enumerate}
%See the front matter of this document for examples. 
%You are recommended to conform your choice to the journal you 
%are submitting to.
\subsection{The learning management system}
Canvas is the Learning Management System (hereinafter called learning platform or LMS) used at [University name blinded for review purposes]. It was implemented during spring 2017. The learning platform includes tools for course content customization depending on the needs and requirements of each course and the teachers' preference. Assignments, Discussions, Modules, Quizzes, and Pages are the components that can be included during the course creation for educative resources provision. Feedback on submissions is provided via SpeedGrader, and the component Grades, is used to manage reporting of grades. 
The common components shared and used by all courses at the university are; i) Modules, ii) Assignments and iii) Grades. Based on that, these three components were selected to analyze the changes in the study patterns in this paper. However, the internal structure of each component may differ in each course; some teachers put everything on Modules, including for example, links for each assignment, quizzes and external resource material. In contrast, other teachers have everything divided among the components; with Modules only containing the course slides and details about material to read, Assignments having the list of tasks with deadlines, and these teachers typically place external resources like discussions and lecture recordings being embedded into their own components (outside of Modules and Assignments). Each teacher has absolute freedom in deciding the course structure in the learning platform. 

The learning platform also allows the embodiment of external resources to support teaching and learning. For example, Echo360 is a lecture capture tool currently used to store, stream and provide access to lecture recordings. This tool is widely used for lecture capture among a large number of educative institutions around Europe and is a tool, embedded in Canvas. Another tool used is Piazza, a web-based threaded forum created and used to support discussion, answer questions and improve communication between teachers and students. Piazza is also a third-party tool, embedded into the interface of Canvas.

\subsection{Data collection and descriptive analysis}
Table \ref{tbl1} shows the five courses at the Computer Science department selected for this case study. A case study is a democratic and participatory approach that aims to create practical knowledge such as lessons learned, principles or implications, too bring together action and reflection through illustrations from a specific case \citep{yin1994case}. Case studies in general can be positivistic \citep{yin2013validity}, interpretive \citep{walsham2006doing}, or critical \citep{carr2003becoming}. The approach in this particular case study is interpretive \citep{walsham2006doing,walsham2015interpreting} primarily grounded in the data derived from the interactions with the students, where we looked for contextual details of the case \citep{miles1994} and study the students' interactions with the learning platform, which outlines the unit of analysis of this paper.

The table also shows the number of students enrolled in each course during the three years of interest: 2018, 2019 and 2020. Other departments such as Psychology and Sport Science had a more challenging change when the pandemic started because, for the first time, the students would have online lectures and needed to interact with the LMS. As those departments did not have a pre-existing practice of using Canvas, any comparison in the online activity with prior years was impossible. Based on that, we took the decision of not including them in this case study.   
The click-stream data was collected via the Canvas API and we captured all the students’ activity enrolled in the five selected courses. All of the selected courses had a high number of students enrolled during the three years of interest. In addition, all had the same teacher for 2019 and 2020 when the pandemic hit, while three of them, TSAM, GHOH, and REIR, had the same teacher for 2018 as well. Furthermore, all five courses were taught during the fall term, during the same 12-week period. 
We consider the fall term of 2020 as the \emph{pandemic term} in our case study.
The data obtained from the Canvas API, includes Timestamp, URL and \verb+user_ID+. 

\begin{table}[width=.9\linewidth,cols=4]
\caption{Number of students enrolled per year}\label{tbl1}
\begin{tabular*}{\tblwidth}{@{} LLLL@{} }
\toprule
Course name & 2018 & 2019 & 2020\\
\midrule
Calculus and Statistics (CAST) & 156 & 131 & 159 \\
Software Requirements and Design (GHOH) & 237 & 319 & 284 \\
Discrete Mathematics 1 (STR1) & 174 & 225 & 226 \\
Computer Networks (TSAM) & 181 & 207 & 194 \\
Algorithms (REIR) & 208 & 207 & 250 \\
\bottomrule
\end{tabular*}
\end{table}

The data was gathered from August to December of each of the three years; this period contains the 12-weeks of lectures, the first (and main) exam period and the grades’ release. The second exam period (the re-take exams) is not included in the data, this period usually occurs early January, but due to the pandemic the second exam period was moved slightly, and a third exam period was established. Due to this imbalance in exam period number between the years, we decided to only include the 12-week period (which includes the original exam period for all years), and exclude the re-take exams altogether from the data. Furthermore, by the time of our research, the data of this third period is not available, which further strengthened our decision to only include the first exam period in this case study. To clarify, the vast majority of students passes during the first exam period. In addition to the click-stream data, the final grades obtained by the students and their assignment submissions were downloaded from the LMS. 

\subsection{Analysis of click-stream data}
The students' activity was filtered using the URL according to each courses’ ID number; additionally, the URL was used to identify the clicks on different components: Modules, Assignments or Grades activity respectively. Timestamp information was used in the generation of the variables: usage ratio during business days and weekends; usage ratio during morning, afternoon, evenings and nights; time-span of the daily interactions with the LMS and course content, and finally, it was used to split the course into three sections to analyze the changes in the activity level during the course. Table \ref{tbl3} displays the definition of the variables created for the analysis. 

\begin{figure}
	\centering
		\includegraphics[scale=.45]{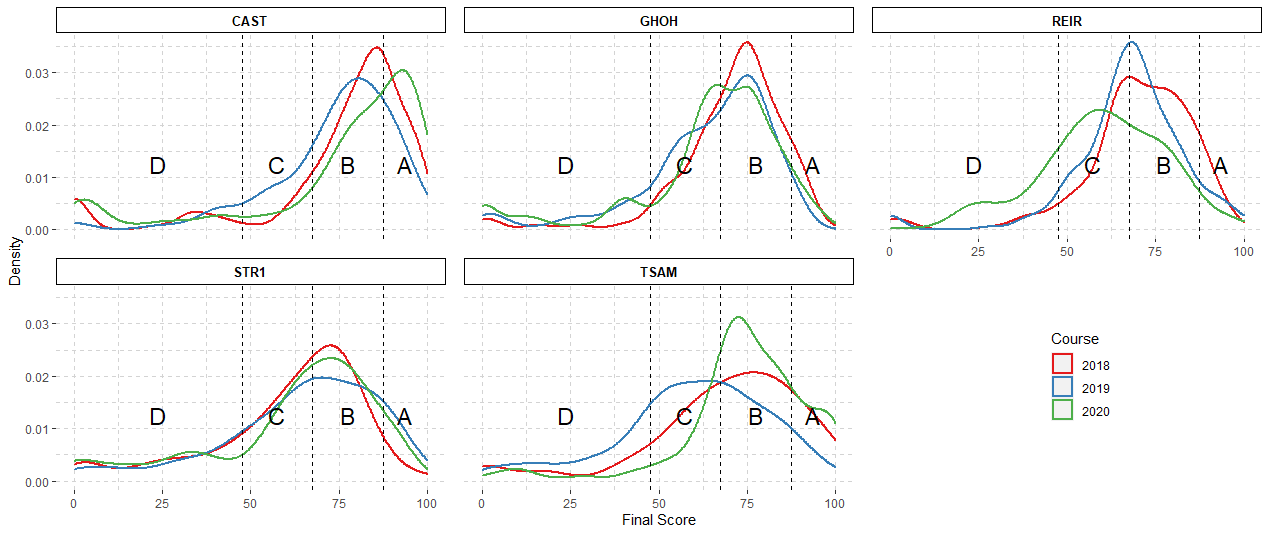}
	\caption{Final Score Densities from 2018 to 2020}
	\label{FIG:1}
\end{figure}

At the university from which the click-stream data was retried, numerical grades are given for each course section. The numerical grade is on a scale between 0-100 where 47.5 outlines a passing grade. For this case study, numerical grades were classified according to the final grade obtained. The numerical grades were replaced, and instead categories were defined from A to D, where A included the students with remarkable performance and D included students who did not pass the course, see Table \ref{tbl3}. Figure \ref{FIG:1} displays the distribution of grades; in these plots it is possible to see how different the courses' grades distribution is among courses, but also among the years of interest. As mentioned before, the courses CAST and STR1 had different teachers in 2018; and the grade distribution of that year is significantly different from the distribution in 2019, rooted in the fact that the teacher and the teaching methods were adopted between those years. 

% Here I would add definitions of all of our variables, i.e. usage ratio, , part of day, part of week, etc. We can do this in a table. 
% We also need to explain what is a module, assignment and grades. Why did we only choose these components?

\begin{table}[width=1\linewidth,cols=3]
\caption{Variable definition}\label{tbl3}
\begin{tabular}{p{3cm}p{2cm}m{9cm}} 
\toprule
\multicolumn{2}{c}{Variable} & Definition \\
\midrule
\multirow{2}{*}{LMS component}  & Modules & Count of URLs containing \verb+"modules"+  \\ \cline{2-3}
& Assignments & Count of URLs containing \verb+"assignments"+\\ \cline{2-3}
& Grades & Count of URLs containing \verb+"grades"+\\ \hline
\multirow{2}{*}{Part of the week}  & Business days & Usage ratio of course content from Mondays to Fridays  \\ \cline{2-3}
& Weekends & Usage ratio of course content in Saturdays and Sundays\\ \hline
\multirow{4}{*}{Part of the day}  & Morning& Usage ratio between 06:00 and 12:00   \\ \cline{2-3}
& Afternoon & Usage ratio between 12:00 and 19:00\\ \cline{2-3}
& Evening & Usage ratio between 19:00 and 24:00\\ \cline{2-3}
& Night & Usage ratio between 00:00 and 06:00\\ \hline
\multirow{2}{*}{Time span}  & LMS & Average daily difference in minutes between the first and the last interaction with the LMS.  \\ \cline{2-3}
 & Course &Average daily difference in minutes between the first and the last interaction with the content of a specific course.\\  \hline 
\multirow{3}{*}{Part of the term}  & Part 1& Activity ratio from week 1 to week 4 of the term \\ \cline{2-3}
& Part 2 & Activity ratio from week 5 to week 8 of the term\\ \cline{2-3}
& Part 2 & Activity ratio from week 9 to week 12 of the term\\ \hline
\multirow{4}{*}{Grade placement}  & A & Students with final grade >= $87.5$ \\ \cline{2-3}
& B & Students with final grade >= $67.5$\\ \cline{2-3}
& C & Students with final grade >= $47.5$\\ \cline{2-3}
& D & Students with final grade < $47.5$. This is a non-passing grade.\\ 
\bottomrule
\end{tabular}
\end{table}

The data on the grades was linked with the behavior data from the LMS. Kruskal Wallis test, One-way, and Two-way ANOVA analysis were used to evaluate significant differences among years, grades, and their interactions with each of the variables created.

\subsection{Assignment submissions}
Additionally to the change in the variables derived from the LMS, assignment submissions are the second student-related component in this case study. In this study, individual assignment submissions reflect the students' personal willingness and motivation to complete and pass the course. However, for group project submissions, it is more difficult to certainly know to what extent a particular student was involved with the assignment. 

To know how the students' willingness and motivation changed during the pandemic, we made assignment submission comparisons between 2019 and 2020; 2018 was not considered because the assignment structure for most of the courses was very different. It is not possible to know if a student who got a zero-grade, submitted the assignment or not, since the learning platform assigns a zero-grade when there was no submission. For our purposes, we define “submission” as all non-zero grades. To make the courses comparable between the years, the percentage of students submitting the assignment was calculated. This comparison was made for all the students, in each course, and for the failed students exclusively to identify the common patterns of those who struggled the most during the term. 

\subsection{Teachers' perspective}
In addition to the aforementioned data, six teachers were interviewed to gather in-depth information about the courses’ structure, the teaching methods used and the decisions made around and during the pandemic restrictions. Furthermore, the objective was also to identify factors and actions related to the pandemic and its limitations on teaching that could have affected the students’ response, interaction level, and performance during the term fall 2020. The CAST course had two teachers during 2019; as the decisions about the course were taken jointly, we decided to interview the two teachers together. The interviews were done following a semi-structured interview guide. As can be seen in Table \ref{tbl2}, there were nine questions: Questions 1 and 2 were about the use of technological resources before 2020, and whether the use changed or not  when the restrictions forced on online teaching; question 3 shed light on the biggest challenges faced by the teacher; question 4 and 5 were related to the course structure, number of assignments, projects, lecture length, etc., before and during the pandemic; questions 6 and 7 asked for the personal attitude of the teachers during the term, and their expectations about the performance of their students during the distance term; finally, questions 8 and 9 were related to their thoughts and expectations about the study and submission patterns of their students. Three interviews were conducted in person, whereas two were conducted through zoom, each interview took around 30 min (but the exact time of each interview varied a few minutes). All interviews were recorded with previous authorization by the teachers interviewed and transcribed verbatim.

% Add a table with the questions 
\begin{table}[width=1\linewidth,cols=2]
\caption{Semi-structured interview guide}\label{tbl2}
\begin{tabular}{p{5cm}p{10cm}} 
\toprule
Topic & Question \\
\midrule
\multirow{2}{*}{Technological resources}  & 1. Tell me about the technological resources you use for your teaching during the term Autumn 2020 (Canvas, Echo, Piazza, etc). Did you use those resources before the pandemic? \\ \cline{2-2}
& 2. How your teaching methods changed due to the pandemic? What did you do different and why? \\ \hline
\multirow{1}{*}{Teaching challenges}  & 3. What was the biggest challenge you faced? \\ \hline
\multirow{2}{*}{Course structure}  & 4. Did you structured your course (number of lectures / assignments / assessment tools) differently? \\ \cline{2-2}
& 5. Which was your rationale about those changes? or to not change anything? \\ \hline
\multirow{2}{*}{Teachers attitude and expectations}  & 6. Tell me about your attitude for the course and your students, did you decide to relax a bit, or to be as strict as in-class pre-pandemic term?\\ \cline{2-2}
& 7. What were your expectations about the performance of your students? \\ \hline
\multirow{2}{*}{Activity level and study patterns}  & 8. Do you think your students had the same study pattern during the term 2020 than the other terms? what changes were you expecting? more or less activity? \\ \cline{2-2}
& 9. How do you think their dropping/leaving the course pattern or attitude changed due to distance learning? \\ 
\bottomrule
\end{tabular}
\end{table}

 The interviews were done during the spring term of 2021, while most of the restrictions were still active. The interview data was analyzed through content analysis \cite{krippendorff2018content}. The data from the interviews helped us to understand particular patterns’ changes in some variables.

\section{Results}

%There are various bibliography styles available. You can select the
%style of your choice in the preamble of this document. These styles are
%Elsevier styles based on standard styles like Harvard and Vancouver.
%Please use Bib\TeX\ to generate your bibliography and include DOIs
%whenever available.

%Here are two sample references: 
%See \citet{Fortunato2010}. Also refer %\citet{Fortunato2010,NewmanGirvan2004}.
%More citations are here \citep{Fortunato2010,Vehlowetal2013}.

\subsection{Click-stream analysis}

\subsubsection{Canvas component analysis}
\begin{figure}
	\centering
		\includegraphics[scale=.35]{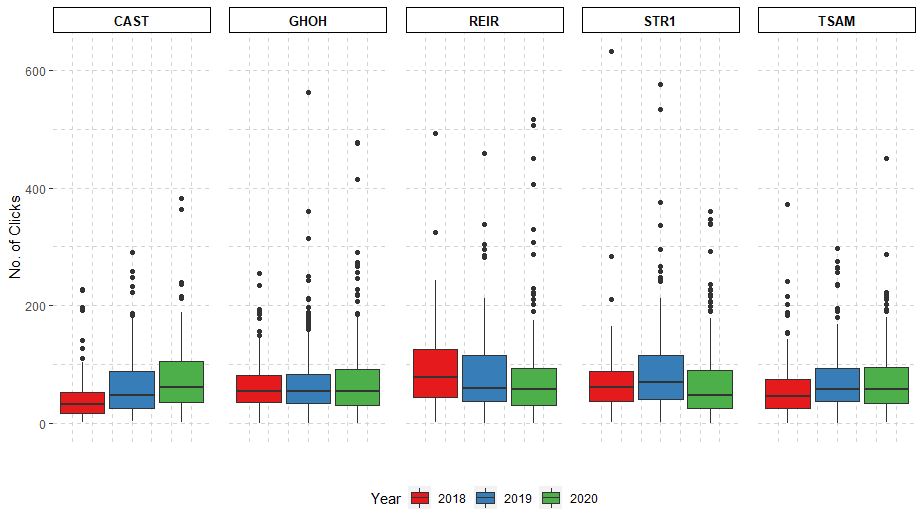}
	\caption{Modules clicks in fall term}
	\label{FIG:2}
\end{figure}

\begin{figure}
	\centering
		\includegraphics[scale=.35]{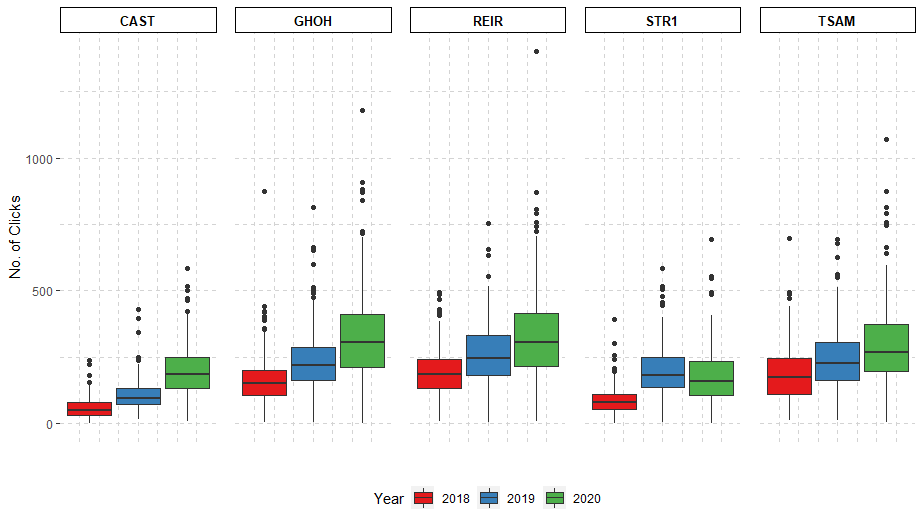}
	\caption{Assignment clicks in fall term}
	\label{FIG:3}
\end{figure}

\begin{figure}
	\centering
		\includegraphics[scale=.35]{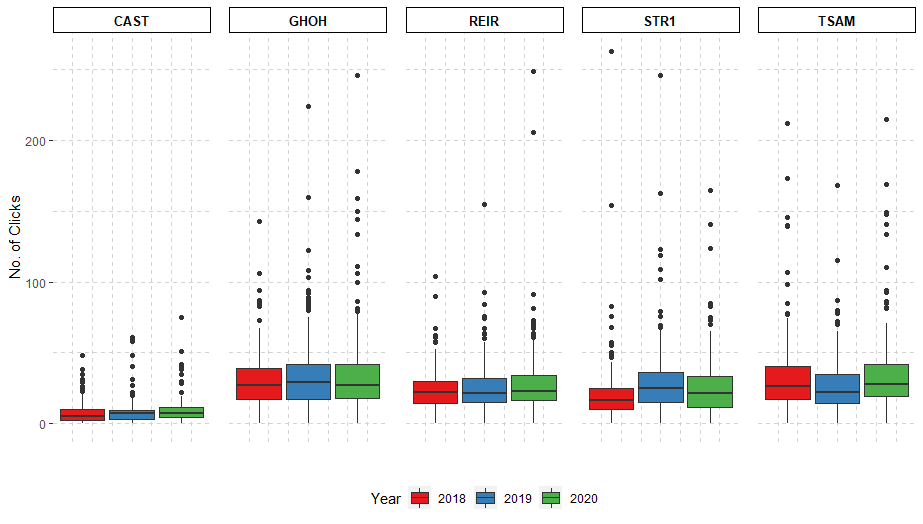}
	\caption{Grades clicks in fall term}
	\label{FIG:4}
\end{figure}

Figures \ref{FIG:2}, \ref{FIG:3} and \ref{FIG:4} compare the students number of clicks in the components Modules, Assignment and Grades respectively. For Modules, there were significant differences in the click distribution for almost all courses and years. Nevertheless, not all the courses had the same behavior from one year to another. These differences are related to the fact that the teacher has complete control of the course’ structure on Canvas. There are not significant differences in the GHOH course, nor the TSAM course from 2019 to 2020. CAST course was the only one with a significant positive increment in the number of clicks in this component. In the assignment component, all the courses showed significant changes for every year. The assignment component, is the component with the most significant changes. Figure \ref{FIG:3} clearly shows the change in teachers in CAST and STR1 courses from 2018 to 2019. Additionally, all the courses except STR1 had an increase in the number of clicks on this component. One important change implemented in the STR1 course from 2019 to 2020 was the creation of an alternative course on Canvas for the students enrolled in the STR1 course. The students were enrolled in both courses, and the interaction was split among the two. In contrast, the grades component is the one which has the least significant changes. Like the modules component, the grades component is up to the teachers’ control; and the decision about the use of it, directly affects the number of clicks on this component. For example, the CAST course distribution is highly different from the other courses; one explanation for this difference in behavior, is that the teachers in this course decided not to release the grades in Canvas, but in the administrative reporting system instead.
Note, that the fewer number of click in 2018 could be party explained by the fact the Canvas was not fully implemented by then.

\begin{figure}
	\centering
		\includegraphics[scale=.5]{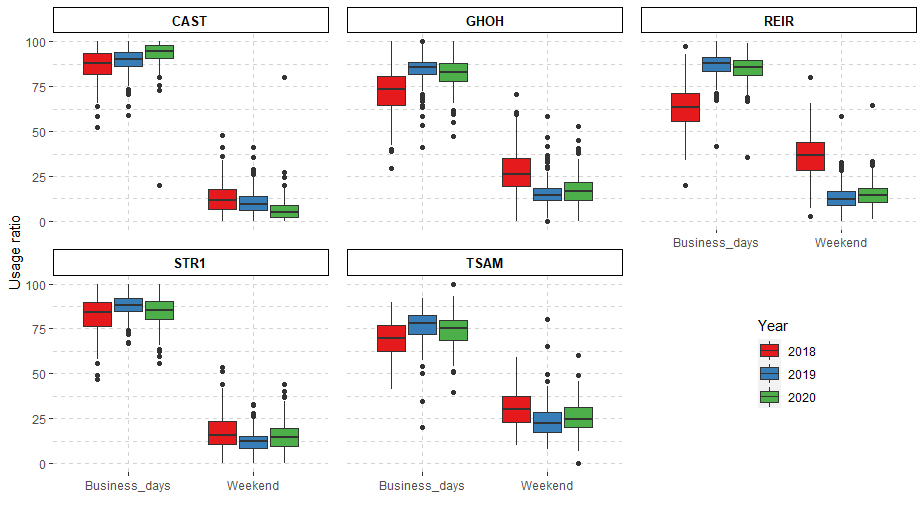}
	\caption{Part of the week usage ratio}
	\label{FIG:5}
\end{figure}

\begin{figure}
	\centering
		\includegraphics[scale=.5]{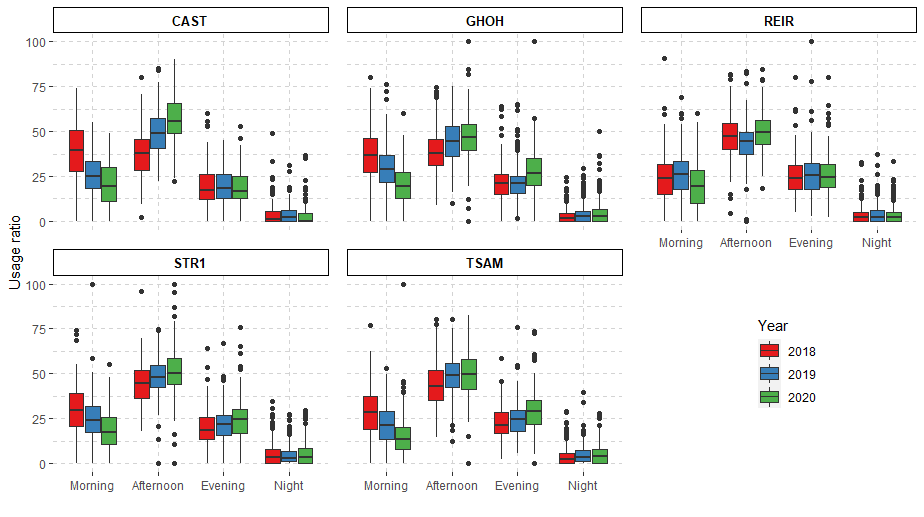}
	\caption{Time of day usage ratio}
	\label{FIG:6}
\end{figure}

Regarding to the ratio usage during weekends and business days, the proportion of time spent by students interacting with the LMS is higher during business days in comparison to the weekend. Figure \ref{FIG:5} displays the usage ratio for the five courses included in the case study, all had significant differences through the years with the same pattern: in 2019, the proportion of time spent during business days increased in all the courses. In contrast, during fall 2020, this ratio had a significant decrease for all courses, except for CAST; which was the only course with an increment in the activity usage during business days. Related to the time of day, figure \ref{FIG:6} shows the proportion spent in different moments of the day. The students interacted more during mornings and afternoons, and those were the times with the most significant changes in usage ratio. In general, during the pandemic the proportion of time during mornings decreased while the ratio in the afternoon increased in a significant amount in all courses. During evenings, increases in the ratio were found in first, and third-year courses. Regarding nights, only CAST course showed a significant decrease in night activity during the pandemic. 

\subsubsection{Grade placement}
\begin{figure}
	\centering
		\includegraphics[scale=.45]{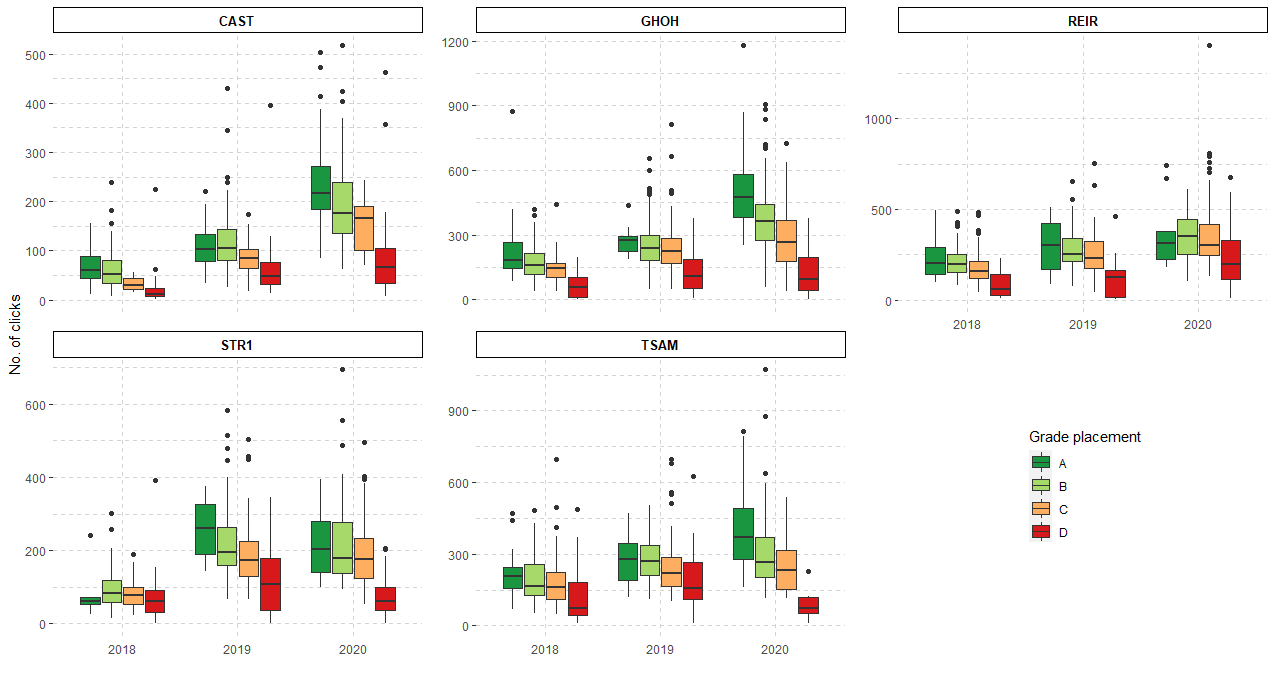}
	\caption{Assignment clicks by grade placement}
	\label{FIG:7}
\end{figure}

\begin{figure}
	\centering
		\includegraphics[scale=.45]{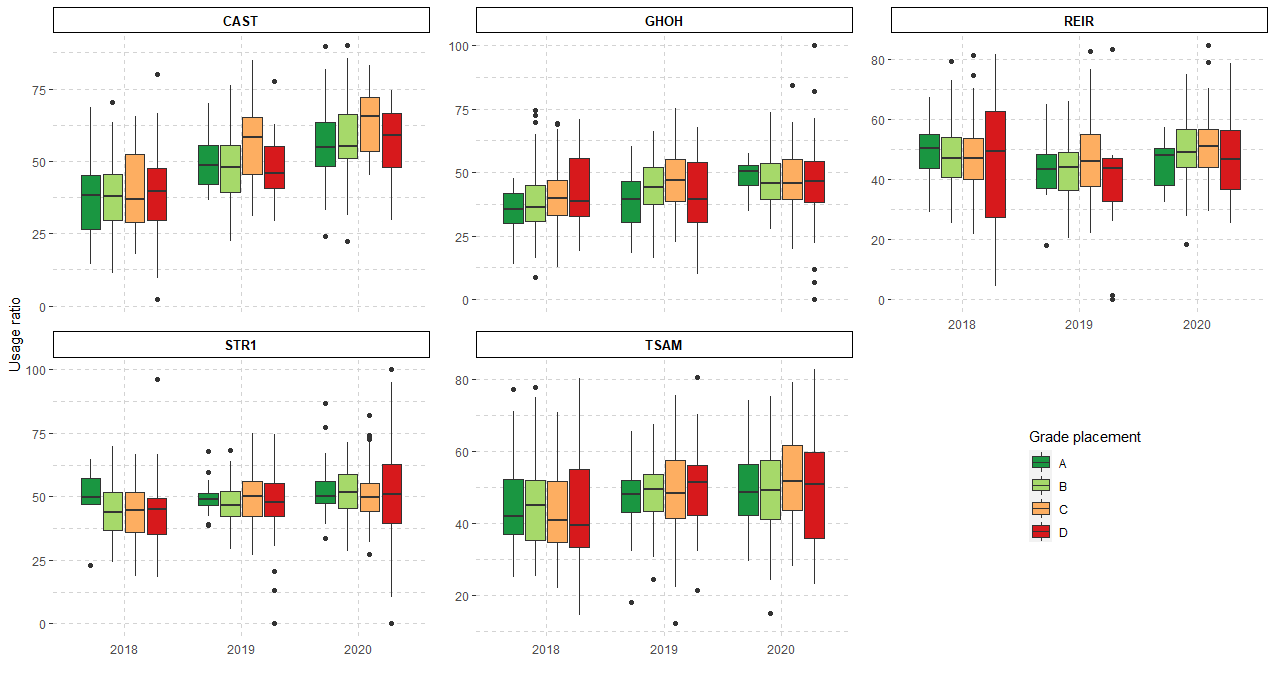}
	\caption{Afternoon usage ratio by grade placement}
	\label{FIG:8}
\end{figure}

In addition to the changes in usage of the LMS, we analyzed the differences in those variables when the students are classified by grade placement to identify changes in the behavior and usage of it. Figure \ref{FIG:7} shows the interaction between grades and assignments' clicks. Statistical analysis showed significant differences in the distribution of clicks according to grade placement, being the grade D the one with differences for all courses. Independent of the teacher and year, students with D grades are significantly less engaged with the assignment component in comparison to those students who passed the course. This pattern is similar in modules and grades clicks. In contrast, our analysis did not find any pattern in the ratio usage for different days of the week or times during the day. Figure \ref{FIG:8} displays the interaction between afternoon usage ratio and grade placement. Year is the only significant variable that influences the usage ratio in these courses, and there is not any difference in the time and day when the students use the LMS, and the grade they get.  

\subsubsection{Average daily time span}
\begin{figure}
	\centering
		\includegraphics[scale=.45]{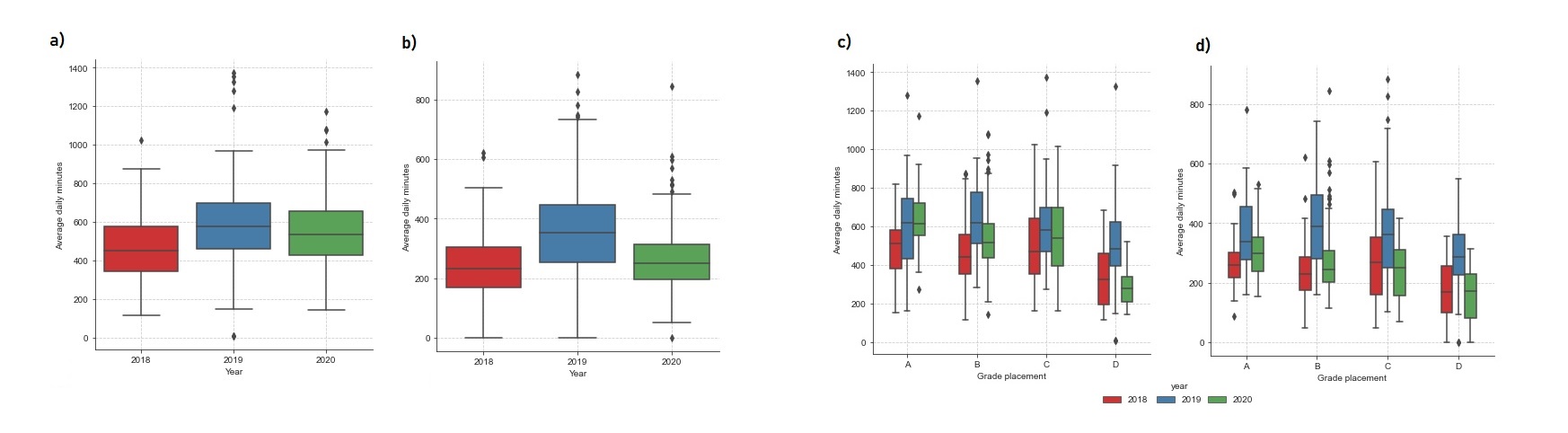}
	\caption{Daily time spent interacting with the LMS and couse content for TSAM students}
	\label{FIG:13}
\end{figure}

In this section, we created a variable for the average daily minutes as the difference between the first and the last interaction in the same day. This variable was created for both the interactions within the course content and with the LMS Canvas, independently of the course. Plots a) and b) in Figure \ref{FIG:13} show the distribution of that variable in TSAM course in general, and for the course interactions respectively. In the first plot, it can be seen that the length of the interaction with the LMS per day did not change from 2019 to 2020, but the minutes spent interacting with the course content were fewer than the time span in 2019. The students did not increase the time interacting with the learning platform, but their interactions show that they were more focused while engaging and studying with the content of each course they were enrolled in. This pattern is present in all the courses included in this case study. Considering this pattern jointly with the results previously presented, it can be concluded that the activity was more intense, with more clicks, highly concentrated in a smaller time span during the day.
Plots c) and d) split the results in a) and b) by grade placement. While in plot d) the reduction in the time span is similar for all the grade placements, in plot c) is remarkable the difference in grade placement D. This difference indicates that students who failed TSAM course not only spent less time interacting with the course content, but they also spent less time interacting with the LMS in general. This behavior is particularly visible for TSAM course, and is not shared by any other course in this case study. 

\subsubsection{Temporal changes}
\begin{figure}
	\centering
		\includegraphics[scale=.45]{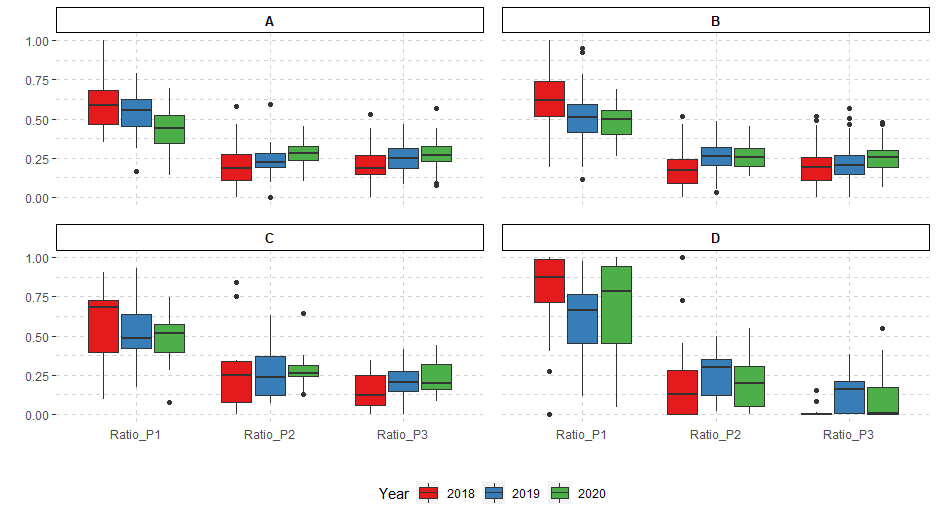}
	\caption{CAST usage ratio by grade placement}
	\label{FIG:9}
\end{figure}

\begin{figure}
	\centering
		\includegraphics[scale=.45]{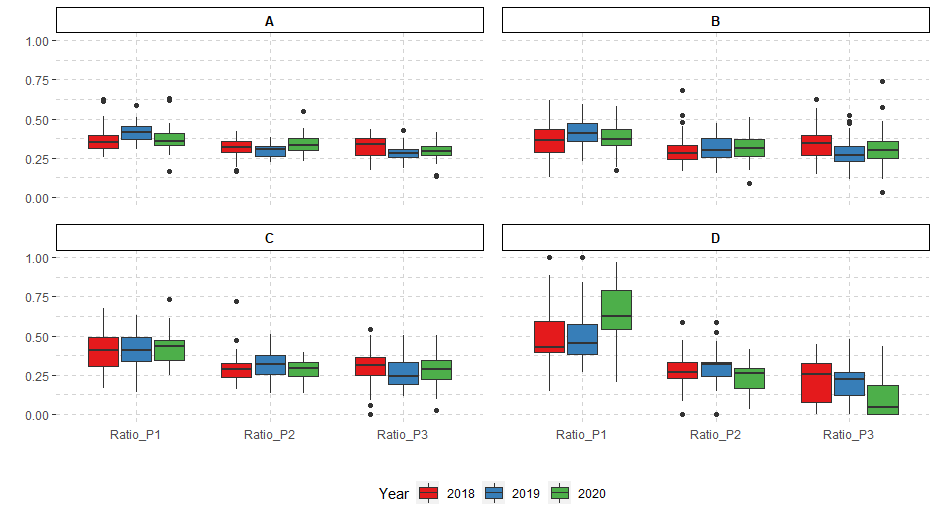}
	\caption{TSAM usage ratio by grade placement}
	\label{FIG:10}
\end{figure}

In this subsection, we take a look at the \emph{part-of-term} variable  which evenly divides the 12-week course within three parts and compare the usage ratio in each of those parts. The split had a significant effect on all the courses. Figures \ref{FIG:9} and \ref{FIG:10} show the ratio usage distribution of CAST and TSAM courses for each grade placement respectively. Among the patterns observed on the usage ratio distribution, the students who got a D grade and failed the course had a distinguishable usage behavior, with the higher percentage of their usage concentrated in the first four weeks of the course. The usage after the first part of the course shows a sudden drop for the last eight weeks; this indicates that those students stopped using the learning platform. In contrast, for the other three grade placements of students with a passing grade, the decrease along the course was smoother, indicating that even when the students were interacting less, they still dedicated a significant amount of time. Except for TSAM course, no significant changes in the distribution of time of failed students during the term were identified; nevertheless, one significant change was identified in TSAM course. The ratio of usage during the first four weeks by the D grade students, was higher than it was in the previous years. 


\subsection{Assignment submission analysis}
\subsubsection{Percentage of submission}
\begin{figure}
	\centering
		\includegraphics[scale=.55]{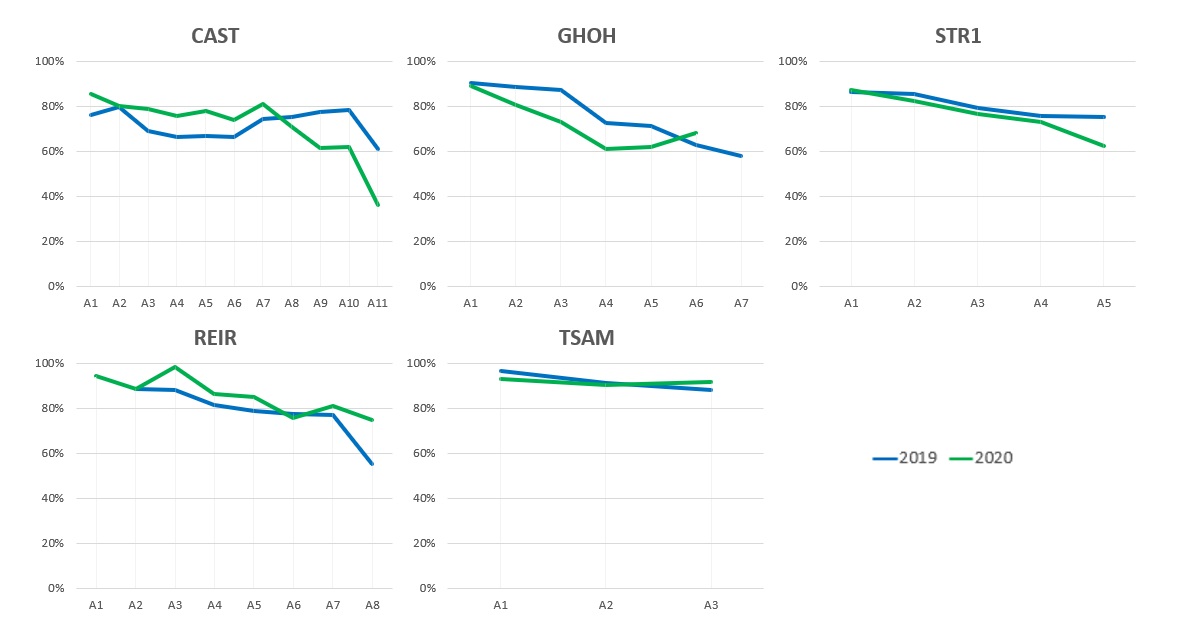}
	\caption{Percentage of submission}
	\label{FIG:11}
\end{figure}

\begin{figure}
	\centering
		\includegraphics[scale=.55]{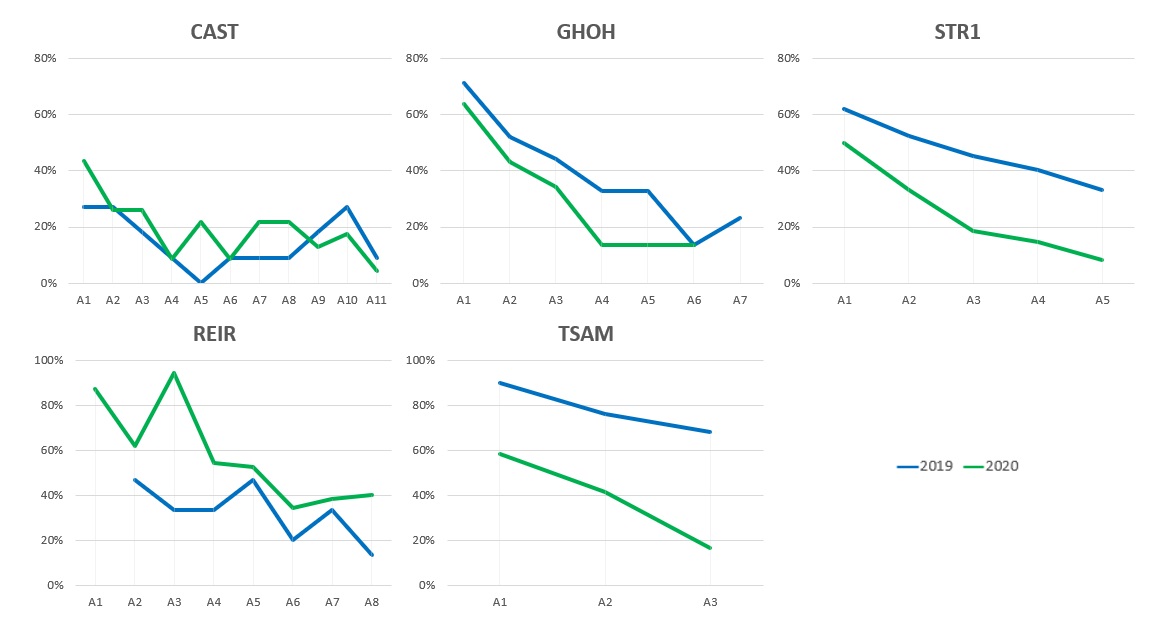}
	\caption{Failed students' percentage of submission}
	\label{FIG:12}
\end{figure}

From the interview with the teachers, we confirmed that all the courses, except for TSAM, had similar assignment structure in 2019 and 2020. Fig \ref{FIG:11} and \ref{FIG:12} display the percentage of submission for all students and failed students exclusively. Valuable differences are observed in the percentage of submission' comparison; the following patterns were identified in the analysis. 

The students in their first term at the university, struggled to accomplish the submission deadlines. For GHOH and STR1 courses, the plots show a lower percentage of students submitting their assignments during 2020. As mentioned by one of the teachers during the interview, one concern was the ability of the first year students to easily adapt to all the systems within the teaching platforms at the university. 
Fig \ref{FIG:1} shows that these courses were also those with similar grade distribution in all years. 

Before the pandemic, the CAST course had weekly individual assignments to be finished during the lecture session without the use of the book, a situation that could not be controlled during the year 2020 when the students were working at home. The assessment of individual assignments in this course was the same in the three years, only the higher grades contributed to the final mark while the lower three or four grades were not considered. Fig \ref{FIG:11} shows that for the first eight weeks, the percentage of submissions was higher than the year 2019, presumably because the students solved the assignments with the use of the book. In contrast, the percentage of submissions for the last four assignments was considerably lower than the percentage in the previous year. In addition, this period happens to cover the chapters in the book about statistics. Each year the students of CAST course have problems to address the requirements of this topic as a combination between complexity and workload at the end of the course. This pattern is also present in the failed students' submissions plot. Still, the percentage of submissions is significantly lower than the average percentage of the entire group of students, indicating less engagement and willingness since the beginning of the course. 

Noteworthy, TSAM course shows the difference in the percentage of submission between the failed students and the general average submission percentage despite it not having the same assessment criteria as in the year 2019. This course had the particular change of a series of weekly quizzes added to the assessment, those quizzes led to an increase in the activity and engagement of the students. However, students who did not pass the course did not contribute to the increased activity. This is evident in the low submission percentage, in the change in the usage ratio, and also in the decrease of the average minutes of interaction.

Finally, for REIR course, both Fig \ref{FIG:11} and Fig \ref{FIG:12} shows that the percentage submission during 2020 was higher than in 2019. The difference is bigger in the first four assignments. In the interview with the teacher, we learned that at the beginning of the course, the assignments were changed from pair projects to group projects. This change was implemented with the intention of improving the interaction of the students with their peers. However, after the first half of the course, that new rule was reversed from groups to two students, because there were numerous problems with the group arrangements. This course clearly shows that group assignments increment the percentage of submission, but that increment cannot be translated directly to a passing grade placement.

%\subsubsection{Submission of failed students}
\subsection{Teachers' perspective}
\subsubsection{Technological resources}
As stated before, the courses selected had previously implemented the usage of the LMS into the everyday learning strategy. During the interview, the teachers were asked about the resources used before and during the pandemic. The five courses had lecture recordings available for the students in the learning platform, both in 2018 and 2019. In four out of five courses, the live lectures were recorded and stored in the Echo360, which is embedded into the learning platform, and only REIR course streamed the lectures live. In addition, there was one course which used old recordings from a previous teacher; despite being outdated for the current course syllabus, the teachers found the content helpful for the students. The tool used by the teachers for storage and facilitate the lecture capture, both for streaming and for pre-recorded lectures, was the aforementioned Echo360. During the autumn term of 2020, all the lectures were recorded and distributed to the students directly on Echo360. REIR lectures were streamed, as in the previous years; in addition, only one more teacher attempted to stream using the LMS; nevertheless, after a few sessions, that teacher changed to pre-recorded lectures, due to both technical problems with the streaming, and also because the majority of the students were not attending the streamed lectures, but instead preferred to watch the recording at a different and more convenient time for them. Although not using the tool Echo360 for the first time, only REIR teacher declared to be using more variety of the features available, like multiple choice questions embedded in the slide deck to trigger engagement.  

Among other technological resources used in these courses during the pandemic, Zoom was used to hold live problem solving sessions. Zoom is also embedded into Canvas, and is a third-party tool as well. In addition, all teachers made use of Piazza tool (the threaded forum), whether for solve questions or to assist during home exams. Other tools such as My Open Math, Web works, Digi-exam, Mooshak and EMT were used but not shared among all the courses as they addressed specific needs of each course. 

\subsubsection{Teaching challenges}
Among the teaching challenges faced by the teachers, the lack of direct contact with the students was identified by most of them as the main challenge during the distance learning term. The lack of interaction with the students complicated the identification of those topics that the students found difficult to understand. The teachers elaborated that knowing what the students had difficulties with was much easier in previous years with live lectures in class, as the students' facial expressions were helpful to identify whether or not they were connecting with the material. Low involvement and motivation were also mentioned by the teachers, for instance, those who ran live sessions to solve questions found poor student engagement and participation. It was usual for the students to attend, but the students were often silent and did not take advantage of the time with the teacher. 
Another challenge for the teachers was the increased workload due to: (1) The need to record everything in advance to make it available for the students, (2) Making clarifications or corrections to the material delivered or previously recorded was complicated because the videos must be edited which can be extremely time-consuming, (3) Adjustments needed in the LMS, take much time regardless of the magnitude of the adjustment, (4) Grading needs, as the students were not always able to arrange group works, a large portion of assignments were delivered individually, increasing the number of submissions, and (5) Lack of sessions having all the students at the same time, implied answering questions repeatedly for many students or in more than one tool.  
Finally, cheating on exams and selecting essential course content to keep, were also mentioned by one teacher.
\subsubsection{Course structure}
The structure remained mostly similar in all the courses. Most of the teachers decided to keep the syllabus as it was before the pandemic. For those who decided to modify it, like GHOH teacher, the rationale was to select those topics with less impact on the subsequent terms and those that the students could learn by themselves when needed to. Regarding the number of assignments, there were small additions in some courses: CAST included one video component assignment where the students recorded a video explaining the solution for one exercise; STR1 added one optional assignment and also included one weekly summary of activities; the most significant change in REIR course was related to the assignments, which were in pairs before the pandemic, but to facilitate peer interaction were changed to group assignments with up to five students in each group, participation was also included as part in the final mark. In contrast, due to pandemic restrictions, GHOH and TSAM teachers allowed more individual submissions to assignments that were group submissions the previous year. TSAM was the course with the most changes; weekly quizzes were included for the first time as part of the assessment while the final exam was in multiple-choice question format. Those modifications in the courses changed the weights of the assessment components in a small proportion. 
\subsubsection{Teachers' attitude and expectations}
In this section, we elaborate on the teachers' views about their attitude with the students and also the expectations they had about their performance. Four out of six teachers declared to have a less strict criteria with the students. The remaining two teachers stated that they were as strict as ever, with some exceptions for extraordinary situations understanding the particular psychological issues and personal problems that some students may face. Some examples of relaxation provided were: i) being benevolent with the deadline of the submissions, ii) being extremely detailed and specific in the explanations and videos, and iii) simplifying the classes and assignments. 

Regarding the performance expectations, the opinions were split: CAST teachers were expecting bad performance due to the situation, restrictions and technical difficulties; similarly, TSAM teacher said not to expect a better performance but felt that the multiple-choice questions format of the final exam helped the students to perform better; STR1 teacher expected that more students were able to finish the course successfully, and have better grades compared to the previous years; GHOH teacher expected that the students were able to get the essentials of the course and to have at least a similar performance than in previous years; finally, REIR teacher expected a similar performance than in previous years considering he ran his course as usual as the situation allowed it. 

In the same way opinions and expectations were divided, so were the results: notwithstanding the lower percentage of assignment submissions, STR1 and GHOH courses do not show a statistical significant change in the grades distribution between 2019 and 2020. Both courses are compulsory courses for first-year students, so these results may indicate these students can overcome difficulties and important changes in their education. The other three courses showed significant changes in the distribution of the grades between 2019 and 2020. In contrast to SRT1 and GHOH courses; even with a higher percentage of submission due to group assignments at the beginning of the term, REIR course showed a higher percentage of students with lower grades. This result shows the organic effect of the pandemic on the students' performance when fewer changes are implemented: the material provided, the lectures, assignments, and resources are the same as before. The students in this group were in their second year of study, and they were used to the learning resources used by the teachers, and even with the possibility to work in groups, interact with their peers, and get points for participation, their performance showed they did not manage to adapt to those changes successfully. The last two courses, CAST and TSAM also showed significant changes in the distribution of grades, having a higher percentage of students allocated in higher grade placements. The possible explanation for those variations may correspond to the open-book assignments and the multiple-choice questions exams.  
\subsubsection{Activity level and study patterns}
In this section, we relate to the teachers' view about their expected changes in the student activity related to their course. The general expectation elaborated in the interviews was that the students would be less active compared with the previous years. The teachers that expected more activity and involvement were TSAM ad REIR teachers; due to the weekly quizzes and participation component implemented, they expected more activity and interaction with the recorded lectures. 

In relation to the decrease in activity level and dropping rate, all teachers agreed on expecting a drop-out rate similar to the previous years. Some of them mentioned that being at home, could facilitate the students to persist in the course up to the final exam date. Our results do not show evidence that supports that hypothesis.

\section{Discussion and Conclusion}

This study investigates the impact of the pandemic of COVID-19 on study pattern behavior in higher education in an extensive manner. To this aim, we analyzed three components involved in the changes in study patterns, through a case study which relies on interaction with the LMS through click-stream data, assignment submission and teachers' perspective through semi-structured interviews. This combination of a data-driven and qualitative approach allows us to assess the activity of all students, and reflect on the view of the teachers, in the selected courses, avoiding the limitations present in other approaches like surveys and questionnaires. Questionnaires as a data collection method has the benefit of being low-cost and are easily spread. However, their use has several limitations, such as a low response rate and biases in the answers obtained \citep{blumberg_cooper_schindler_2011}. The first form of bias is introduced due to the possibility of answering the survey. During the pandemic, those students with limited access to technology resources and internet connection may not be able to answer the survey or decide not to do so and prioritize the study time over the survey. As a second form of bias, subjectivity when reporting changes in study time may be present due to self-perception and personal feelings \citep{stockemer_2019, fowler2013survey}. A third limitation of using anonymous surveys and questionnaires for a purpose like the one explored in this particular case study, is the impossibility of linking the grades obtained with the pattern study changes identified. Furthermore, the results that we have seen so far in relation to the impact of the pandemic on study pattern behavior, rely on questionnaires \cite[cf.][]{Aucejo20,benito2021}, which shows that there is a gap in the literature regarding data-driven approaches that explore engagement of students, through alternative approaches. That gap is what our paper contributes to.

Data collected from the LMS Canvas was used to create a set of variables related to the time distribution and study pattern: activity time span, ratio usage of part of the day and week, activity level in different parts of the term, interaction amount with the LMS components, submission rates and grade placements. Six teachers were interviewed about their expectations, challenges faced, and decisions made during the term to meet the teaching requirements of the ERT period. Our study included the activity of five compulsory courses at the Computer Science department at [University name removed for review purposes]. The data was gathered from 2018 to 2020 to properly assess the changes in teachers, assessment methods, and pandemic conditions of teaching.

The analysis revealed significant changes in the interaction levels with the LMS, the highest increment was in assignments module. Regardless of the assessment method, number of assignments, and number of lectures all courses had a significant increment in this component. Similar conclusions about changes in the usage of an online platform during the pandemic were made by \citet{reigal2020}. Whereas all courses showed a significant change in the distribution of activity during the day, with less activity during the morning and more during afternoons; there were not common changes in the distribution of activity between business days and weekends, with only one course with an increment during business days. \citet{Noviyanti20} results showed changes in the sleep pattern before and during the pandemic, when the students declared waking up at later hours. 

Activity level and interaction with the components in the LMS had an impact on the grade placement obtained by the student. Despite not all courses and grade placement experienced the same impact on activity level, important changes in failed students were found in courses where the teachers declared that they made an effort to keep the course and their attitudes towards the students the same during the pandemic term as it had been in previous years. It was also found that the study time during the pandemic was more intense and more focused. Due to changes in the distribution of time among the courses that the students were enrolled in; they had more activity in each course, in a smaller time span. Nevertheless, no changes in the overall time interacting with the LMS were found during the pandemic.  These conclusions can be contrasted with those obtained by \citet{Aucejo20} about changes in study hours and with \citet{Noviyanti20} conclusions about changes in intensity level during the pandemic in high school students.
Regarding to activity level changes in different moments during the term, the proportion of activity does not show significant changes compared with the previous years. Grade placements A, B, and C, had more even activity levels than grade placement D; whose students have the majority of their activity concentrated in the first four weeks of the term. 

Differences related to the year of teaching in each course were also found. In agreement with authors like \citet{benito2021} and \citet{zhang2021measuring}; first-year courses showed that during the pandemic the students struggled to accomplish the submission deadlines, with submission rates lower than the rates in 2019. Despite that, those students demonstrated being able to adapt to the new environment, systems, and requirements at university; the distribution of grades was similar during the three years analyzed in this study. 
Second-year courses showed an increment in the submission rate during the pandemic. That increment is related to the possibility to solve the assignments using the book and other resources, or with the possibility to work in groups instead of alone. Nevertheless, the increment in submission rate cannot be translated directly to an increment in the proportion of failed students, the distribution of grades during the pandemic was significantly different from the distribution in previous years.
In third-year courses, we found that those students who failed had important changes in their study patterns compared with the failed students in previous years. Their submission rate during the pandemic was significantly lower than the previous year; their time span interacting with the LMS and with the course content was smaller than the time span in 2019; and the proportion of activity during the first part of the term was greater than the proportion in previous years, indicating that in 2020 the students gave up earlier during the term. 

Teachers adapted their teaching methods to the existing technological resources that were used before the pandemic. However, despite all of them having previous experience with those resources, an increase in the workload, lack of interaction, less student motivation and involvement were common denominators in the teaching experience during the pandemic. The teachers kept the same course structure as the previous years. The changes were implemented to support the students' interaction with their peers and keep them interested in the course content. Nevertheless, the pandemic had important limitations for group working and the students preferred to work alone instead of working in groups. Related results on workload have been found in research studies led by \citet{benito2021}. Several authors also concluded motivation and engagement levels were affected negatively by the pandemic \citep{code_ralph_forde_2020, Survey20, benito2021, Aucejo20} 

This case study has three main limitations. Firstly, the data collected with Canvas API does not include activity using mobile phones or tablets. Even though it is not possible to easily access that data to include it as part of this study, most of the student activity before and during the pandemic was focused on the assignments component. Due to the nature of that component and its relation with submission commitment, our results effectively reflect the overall study pattern followed by the students. The second limitation is that the lectures were recorded, stored and provided using the third-party tool Echo360. Data collection through Canvas API does not allow access or to analyze the interactions and usage of the posted videos in that platform. Deep analysis of the use and interaction with the recorded lectures was not part of this study; indeed, adding Echo360 activity to the activity of the components included in this study may lead to confusion in the study patterns identified. Nevertheless, this new component, whose use was not mandatory before the pandemic, should be independently analyzed. In the future, more research will be conducted to investigate the usage of external tools like Echo360 and Piazza, to incorporate them into this research to enforce the pattern analysis performed in this study. The third limitation is that personal problems and barriers faced by the students to access the educational resources provided are not included in the analysis. Despite the distribution of satisfaction surveys among the academic community, those surveys were answered anonymously, keeping us from linking the results with the interaction and grades. This limitation is the hardest to address. Some attempts with resources that allow access to \verb+student_ID+ have been made during 2021, but a low response rate is still present. In the future, non-invasive tools that allow us to collect data about students' feelings, motivation and personal issues should be implemented before we were able to incorporate this component into our research. The findings can be used to support teachers in the creation of teaching and learning material that adequately addresses the particular needs of students depending on their year of study, but also taking into consideration the changes in pattern study, distribution of time and activity through the term. 

More than a year has passed since the pandemic started. Education, as many other sectors, has managed to adapt to the requirements and barriers implied by the impossibility to teach students face-to-face as it was done before. Its effects are still present and will be here for a while, students and teachers around the world are still working hard and doing their best from home. In this paper, we showed how the COVID-19 pandemic and those changes implied by it, impacted study patterns of students in higher education. More specifically, we had the objective to investigate the first insights of education's new normality to support the adaptation of higher education to it, to support students addressing their particular needs, and also to support the academic community to understand more deeply the effects of the pandemic in the study patterns of undergraduate students and the changes in the interaction with the teaching and learning material.  
\bibliographystyle{cas-model2-names}

\bibliography{cas-refs}




\end{document}